\documentclass[aps,prx,twocolumn,10pt,superscriptaddress,amsmath,amssymb,showpacs]{revtex4-1}
\usepackage{graphicx}
\usepackage{bm}
\usepackage{epsfig}
\usepackage{amssymb}
\usepackage{amsfonts}
\usepackage{braket}
\usepackage{color}

\usepackage{epstopdf}
\usepackage{hyperref}
\usepackage{float}
\usepackage{bibentry}
\graphicspath{{Figures/}}
\long\def\/*#1*/{}

\begin{document}
\title{Divergent bulk photovoltaic effect in Weyl semimetals}
\author{Xu Yang}
\affiliation{Department of Physics,
Boston College, Chestnut Hill, MA 02467, USA}

\author{Kenneth Burch}
\affiliation{Department of Physics,
Boston College, Chestnut Hill, MA 02467, USA}

\author{Ying Ran}
\affiliation{Department of Physics,
Boston College, Chestnut Hill, MA 02467, USA}

\begin{abstract}{ Weyl semimetals (WSM) have been discovered in time-reversal symmetric materials, featuring monopoles of Berry's curvature in momentum space. WSM have been distinguished between Type-I and II where the velocity tilting of the cone in the later ensures a finite area Fermi surface. To date it has not been clear whether the two types results in any qualitatively new phenomena. Here we focus on the shift-current response ($\sigma_{shift}(\omega)$), a second order optical effect generating photocurrents. Surprisingly we find that up to an order unity constant, $\sigma_{shift}(\omega)\sim \frac{e^3}{h^2}\frac{1}{\omega}$ in Type-II WSM, diverging in the low frequency $\omega\rightarrow 0$ limit. This is in stark contrast to the vanishing behavior ($\sigma_{shift}(\omega)\propto \omega$) in Type-I WSM. In addition, in both Type-I and Type-II WSM, a nonzero chemical potential $\mu$ relative to nodes leads to a large peak of shift-current response with a width $\sim |\mu|/\hbar$ and a height $\sim \frac{e^3}{h}\frac{1}{|\mu|}$, the latter diverging in the low doping limit. We show that the origin of these divergences is the singular Berry's connections and the Pauli-blocking mechanism. Similar results hold for the real part of the second harmonic generation, a closely related nonlinear optical response.
}
\end{abstract}
\maketitle

\textit{Introduction---} Recently, Weyl semimetals (WSM) have been discovered in many materials with strong spin-orbit coupling\cite{armitage2017weyl,burkov2011weyl,wan2011topological}. The topology of the electronic band structures gives rise to linear band touching points --- Weyl nodes in momentum space, which are monopoles of the Berry's connection and related to the chiral anomaly in the context of high energy physics\cite{berry1984quantal,adler1969axial,bell1995pcac}. These topological semimetals have been shown to host various exotic properties such as surface Fermi arcs\cite{wan2011topological}, semi-quantized anomalous Hall effect\cite{yang2011quantum}, angle-dependent negative magnetoresistance\cite{nielsen1983adler,son2013chiral,huang2015observation}, novel nonlinear optical effects\cite{de2017quantized,Nagaosa_sciadv.1501524,ishizuka2017momentum,kim2017shift,ma2017direct,PhysRevLett.116.026805,chan2017photocurrents}, and a recent observation of a Colossal Bulk PhotoVoltaic Effect (BPVE)\cite{osterhoudt2017colossal}. The non-linear optical response has received increasing attention as a means to probe the Berry curvature of materials in general. This suggests non-linear optical effects can be used to distinguish between materials with different Fermi surface topologies, a question particularly relevant to WSM. Indeed, shortly after the discovery of the first Type-I WSM material in TaAs,\cite{lv2015experimental,huang2015weyl,hasan2015Sci,hding2015NPhys,yang2015NPhys} it was realized the tilt of velocity of the cone can be severe as to result in finite Fermi surfaces at all doping levels.\cite{soluyanov2015type} Nonetheless a clear distinguishing experimental consequence between these Type-II and their Type-I counterparts has yet to emerge. 

To linear order, the electrons near a Weyl node can be described by the effective two-by-two Hamiltonian: $H^{Weyl}= (\sum_{a}k^a  \alpha^a-\mu) \sigma_0+\sum_{a,b} k^a \beta_{ab}\sigma_b$, where $\mu$ is the chemical potential, $\vec k$ is the momentum relative to the Weyl node, $a/b=x,y,z$, $\sigma_b$ are the Pauli matrices and $\sigma_0$ is the identity matrix. After choosing the frame of the principal axes and a proper basis, the off-diagonal elements in $\beta$ can be eliminated:
\begin{align}
 H^{Weyl}=(\hbar \vec k\cdot \vec v_{t}-\mu)\;\sigma_0+\hbar \sum_a (k^a v^a  \sigma_a).\label{eq:Weyl}
\end{align}
Here apart from the generally anisotropic velocity $\vec v$, another velocity $\vec v_t$ describes the tilting of the bands and breaks the degeneracy between the bands at $\vec k$ and $-\vec k$, with the chemical potential ($\mu$) usually nonzero.

\begin{figure}
    \centering
    \includegraphics[width=0.5\textwidth]{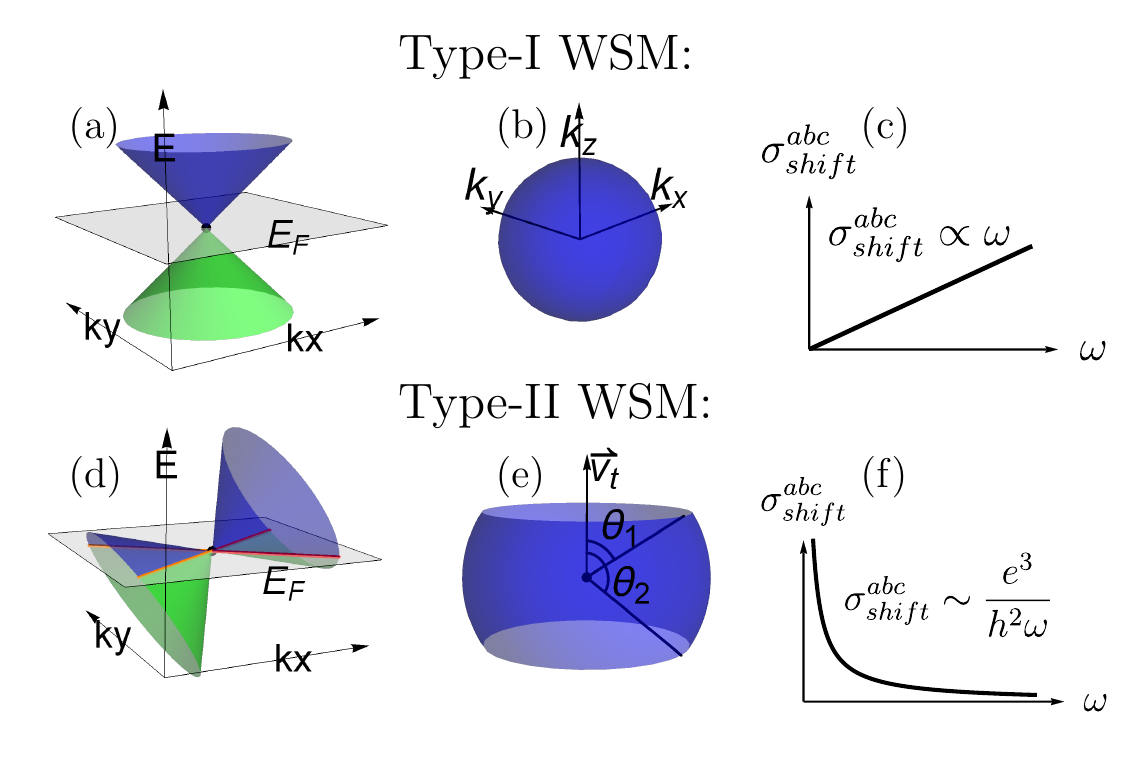}
    \caption{(color online) Considering the $\mu=0$ situation, in (a),(d) we schematically plot the dispersion relations near a Type-I (Type-II) Weyl node. At zero temperature, the momentum space surfaces contributing to $JDOS$ (defined in Eq.(\ref{eq:JDOS})) are qualitatively different in (b) Type-I WSM and (e) Type-II WSM, leading to drastically different scaling behaviors of $\sigma_{shift}$ shown in (c),(f).}
    \label{fig:illustration}
\end{figure}

It is known that either inversion symmetry or time-reversal symmetry needs to be broken in order to realize the Weyl nodes, and the experimentally confirmed Weyl semimetals have been overwhelmingly non-centrosymmetric and time-reversal symmetric, including the (Nb,Ta)(As,P) and W$_{1-x}$Mo$_{x}$Te$_2$ series\cite{lv2015experimental,huang2015weyl,hasan2015Sci,hding2015NPhys,yang2015NPhys,xu2015experimental,soluyanov2015type}. When the tilting velocity is large enough, which is realized in materials like W$_{1-x}$Mo$_{x}$Te$_2$, the system can becomes a Type-II WSM\cite{soluyanov2015type}, since along some directions the two energy bands share the same sign of the velocity. Such a system must have finite size Fermi surfaces  (Fig.\ref{fig:illustration}(d))\footnote{ Consequently it may be more accurate to call it a Type-II Weyl metal instead of a semimetal. However to be consistent with existing literature we follow the existing naming convention.}. The type of a Weyl node is determined by a dimensionless number $W$:
\begin{align}
 W\equiv\sqrt{(v_t^x/v^x)^2+(v_t^y/v^y)^2+(v_t^z/v^z)^2},\label{eq:Type_II_criterion}
\end{align}
and a Type-I(Type-II) Weyl node is realized if $W<1$ ($W>1$). 

As we show later, the bulk photovoltaic effect\cite{sturman1992photovoltaic} with its direct connection to Berry curvature,\cite{Nagaosa_sciadv.1501524,sipe2000second} offers a method to distinguish these two types. It has also attracted significant interest due to its potential applications in renewable energy generation\cite{butler2015ferroelectric} and fast photo-detectors\cite{wang2016review,young2016bulk}. The intrinsic contributions to the BPVE can be expressed as the second order nonlinear photocurrent response:
\begin{align}
 \mathbf{j}^a=\sigma_{\mathbf{2}}^{abc}(\omega) E^b(\omega)E^c(-\omega),
\end{align}
where, $a/b/c=x,y,z$, $\overrightarrow{\mathbf{j}}$ is the DC electric current density, $\vec E(t)=\mbox{Re}[\vec E(\omega)e^{-i\omega t}+\vec E(-\omega)e^{i\omega t}]$ is the electric field of the light. In order to have a nonzero $\sigma_{\mathbf{2}}$, inversion symmetry needs to be broken, which happens to be also a condition to realize WSM. Generally speaking, $\sigma_{\mathbf{2}}^{abc}$ has both intraband and interband contributions. However, for time-reversal symmetric materials with linear polarized light, it turns out that this photocurrent response only has interband contributions, which has been coined the shift-current($\sigma_{\mathbf{2}}^{abc}=\sigma_{shift}^{abc}$) as it results from a change in the center of mass of the electrons upon optical excitation. Perturbation theory within the single-particle framework\cite{von1981theory,aversa1995nonlinear,sipe2000second} gives:
\begin{align}
 \sigma_{shift}^{abc}(\omega)=\frac{2\pi e^3}{\hbar^2}\int\frac{d^3\vec k}{(2\pi)^3}\sum_{n,m}I_{mn}^{abc}[f_{nm}\cdot\delta(\omega_{mn}-\omega)],\label{eq:general_shift}
\end{align}
where $n,m$ label energy bands, $\hbar \omega_{mn}(\vec k)\equiv E_m(\vec k)-E_n(\vec k)$ is the energy difference between the two bands. $f_{nm}(\vec k)=f_n(\vec k)-f_m(\vec k)$ is the difference of the Fermi-Dirac function between the two bands. The gauge invariant quantity $I_{mn}^{abc}(\vec k)\equiv \frac{1}{2}\cdot \mbox{Im}[r_{mn}^b r^c_{nm;a}+r_{mn}^c r_{nm;a}^b]$, where $r^a_{mn}(\vec k)\equiv i\langle u_m(\vec k)|\partial_{k^a}|u_n(\vec k)\rangle$ is nothing but the non-Abelian Berry's connection (with $r^a_{mn;b}(\vec k)$ its generalized derivative: $r^a_{mn;b}\equiv \frac{\partial r_{mn}^a}{\partial k^b}-i[A_m^b(\vec k)-A_n^b{(\vec k)}]r_{mn}^a(\vec k)$, and $A_n^b(\vec k)\equiv i\langle u_n(\vec k)|\partial_{k^b}|u_n(\vec k)\rangle$ the usual intraband Berry's connection).

It has been pointed out that the shift current response is related to the topology of the band structure\cite{Nagaosa_sciadv.1501524}. Indeed the quantities responsible for $\sigma_{shift}$ directly involves the Abelian and non-Abelian Berry's phases. Because Weyl nodes are monopoles of the Abelian Berry's connection, these quantities are expected to be diverging and singular near Weyl nodes, which motivates us to carefully study the resulting nonlinear optical effects. 

Eq.(\ref{eq:general_shift}) has a familiar form of the Fermi's golden rule. Indeed, considering the case of linearly polarized light along the $b$-direction, one has:
\begin{align}
 I^{abb}_{mn}=|r_{nm}^b|^2 R_{nm,b}^a, \mbox{ (no summation on indices)}\label{eq:Iabb}
\end{align}
where the gauge invariant real space displacement $\vec R_{nm,b}(\vec k)$ is the so-called ``shift-vector'', defined as:
\begin{align}
 R_{nm,b}^a(\vec k)\equiv -\frac{\partial\mbox{Arg}[r_{nm}^b]}{\partial k^a}+A^a_n(\vec k)-A^a_m(\vec k),\label{eq:shift_vector}
\end{align}
If one interprets Eq.(\ref{eq:general_shift},\ref{eq:Iabb}) as the Fermi's golden rule, $|r_{nm}^b|^2$ is just the matrix-element factor for the optical absorption. Thus, $e\vec R_{nm,b}(\vec k)$ should be viewed as the dipole moment induced the photo-excited particle-hole pair, giving rise to a rate of change in polarization, i.e., DC photocurrent.

Motivated by photogalvanic applications\cite{butler2015ferroelectric,wang2016review,young2016bulk}, previously the shift-current response has been mainly discussed in the context of insulators\cite{aversa1995nonlinear,sipe2000second}, where the low temperature $\sigma_{shift}$ vanishes when $\hbar \omega$ is below the band gap.  It is convenient to introduce the optical joint density of states including the factor $f_{nm}$ responsible for the Pauli-blocking effect:
\begin{align}
 JDOS(\omega)\equiv\int\frac{d^3\vec k}{(2\pi)^3}\sum_{m,n}[f_{nm}\cdot\delta(\omega_{mn}-\omega)]\label{eq:JDOS}
\end{align}
For most materials, $I^{abc}_{mn}(\vec k)$ is a smooth function of momentum, and according to Eq.(\ref{eq:general_shift}) the response $\sigma_{shift}$ is essentially proportional to $JDOS$. It has been proposed that engineering $JDOS$ in semiconductors may be a route to optimize the BPVE\cite{2017NatCo...814176C}. Although the $JDOS\propto \omega^2$ is small due to the linear dispersion in WSM, recent second harmonic\cite{wu2017NPhys} and photocurrent experiments\cite{ma2017Nat,osterhoudt2017colossal} show that these materials host large nonlinear optical effects at least in the infrared regimes. In this paper we show that the WSM actually feature divergent nonlinear optical responses in the low frequency regime due to the singular Berry's phases near the Weyl nodes.

\begin{figure}
    \centering
    \includegraphics[width=0.4\textwidth]{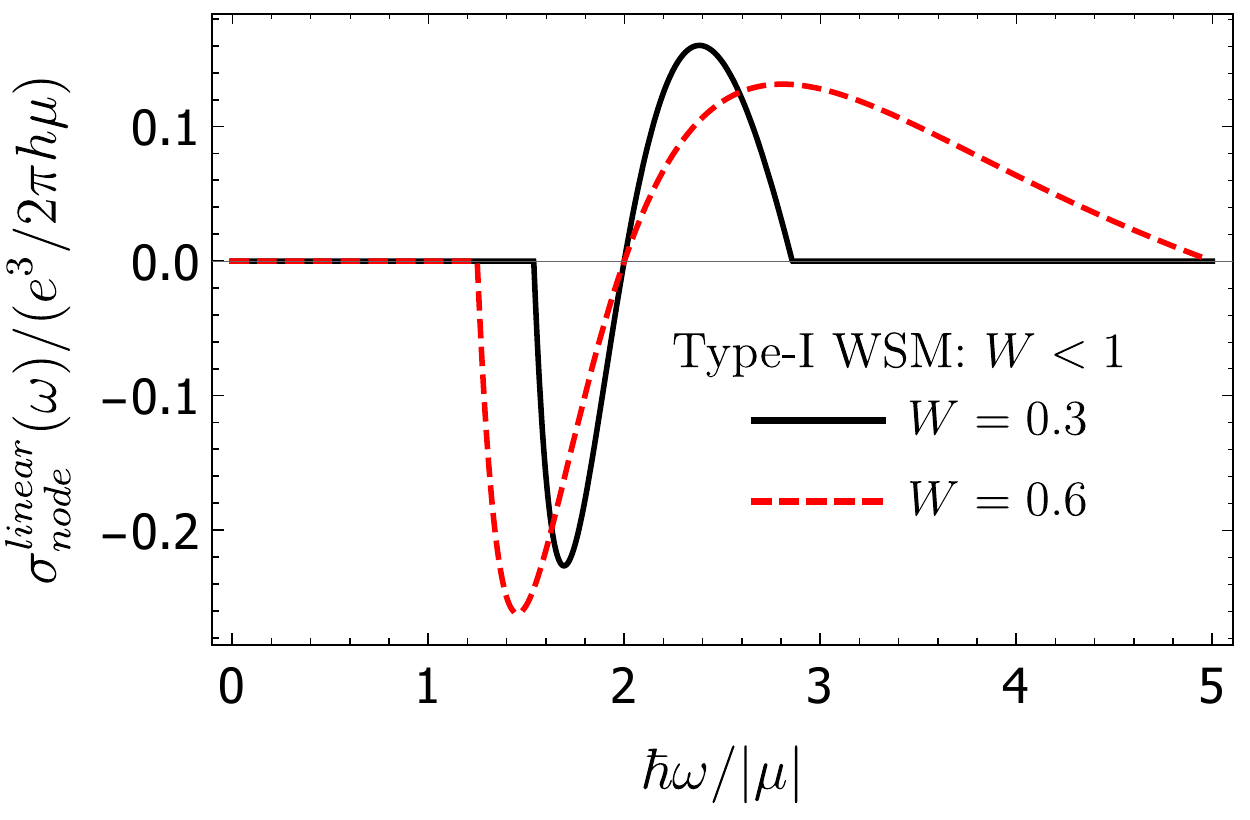}
    \includegraphics[width=0.4\textwidth]{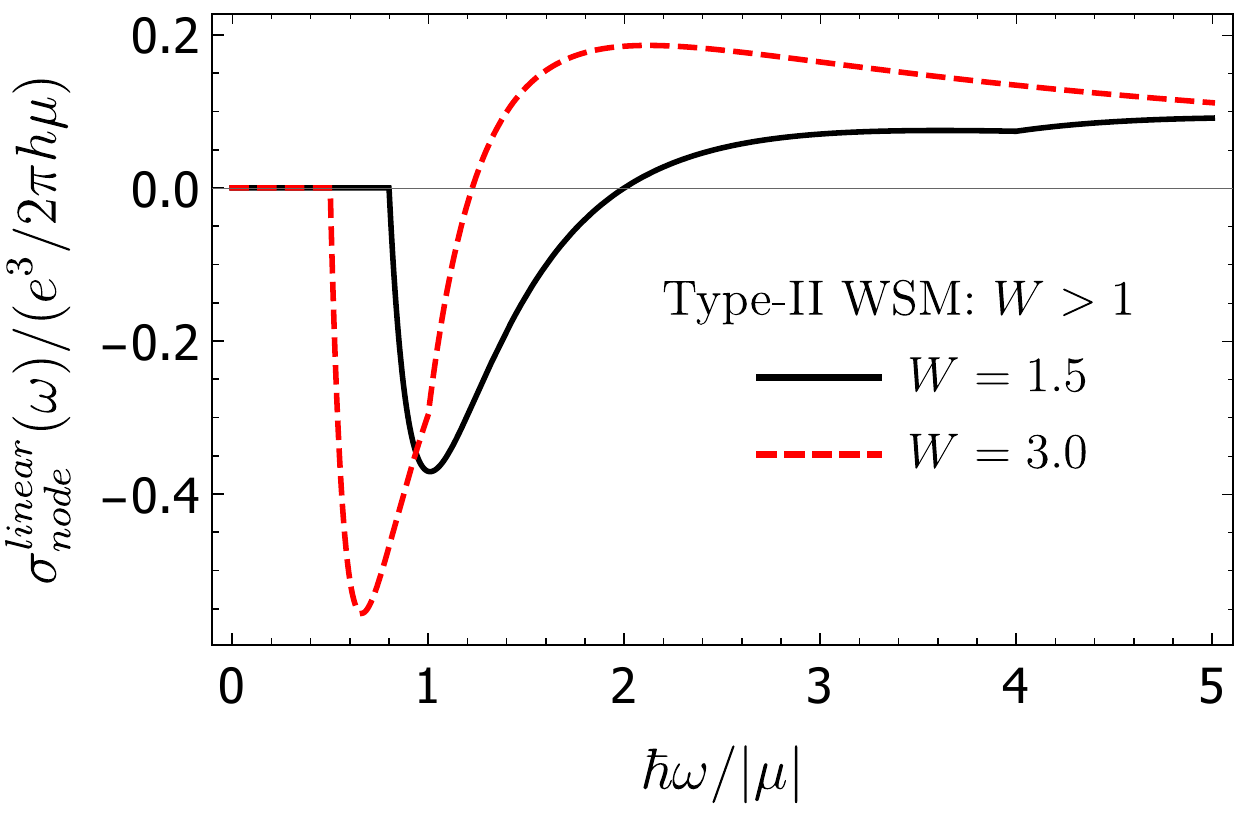}
    \caption{(color online) At $T=0$, the quasi-universal (i.e., $\mu$-independent) line shapes of the doping-induced peaks of $\sigma_{shift}$ in Type-I (top) and Type-II(bottom) WSM based on results Eq.(\ref{eq:sigma_linear_node}) of linearly dispersive nodes. The peak's frequency range has been re-scaled by a $\mu$ factor and its height has been re-scaled by a $1/\mu$ factor, the latter diverges in the low doping limit.}
    \label{fig:line_shapes}
\end{figure}

\textbf{Results:} Our main results are summarized in Fig. \ref{fig:illustration} and Fig. \ref{fig:line_shapes}. For simplicity let us start with the zero doping case $\mu=0$. In the absence of extrinsic scattering processes, simple dimensional analysis shows that up to fundamental constants $\frac{e^3}{h}$, $\sigma_{shift}(\omega)$ is proportional to the inverse of an energy scale, which could involve either an intrinsic energy scale $t$ of the material, the temperature $k_B T$, or the photon energy $\hbar \omega$. 

At zero doping and zero temperature, the linear Weyl equation Eq.(\ref{eq:Weyl}) does not contain an energy scale, therefore $\sigma_{shift}(\omega)$ must be either $\sim\frac{e^3}{h^2}\frac{1}{\omega}$ or vanishing in this linear approximation. The former divergent case is exactly realized in Type-II WSM (Fig.\ref{fig:illustration}(f)). In Type-I WSM, however, we will show that zero doping and zero temperature leads to $\sigma_{shift}(\omega)\sim \frac{e^3}{h}\frac{\hbar\omega}{t^2}$ (Fig.\ref{fig:illustration}(c)), where $t$ is an intrinsic energy scale due to the band-bending (i.e., deviation from linear dispersion) ---typically $\sim$0.1-0.2eV.

Next, we discuss the reason for the drastic difference between Type-I and Type-II WSM, as well as the effect of doping. A careful evaluation of $\sigma_{shift}$ in Eq.(\ref{eq:general_shift}) based on the linear Weyl equation Eq.(\ref{eq:Weyl}) shows that, at $T=0$, a single Weyl node contributes (At $T\neq0$, a general and more complicated analytic result is shown in supplemental material Eq.(\ref{eq:universal_function},\ref{eq:general_sigma_with_universal_function})):
\begin{align}
\sigma^{abc}_{shift}&=\frac{\pi\chi}{8W^2}\cdot\sum_d\big[\epsilon_{dca}\frac{v_t^bv_t^d}{(v^d)^2}+(b\leftrightarrow c)\big]\cdot\sigma^{linear}_{node}(\omega)\label{eq:sigma_line_node}\\
 \;\;\;\;\;\;\;&\mbox{where: }\notag\\
 \sigma^{linear}_{node}&\equiv (\text{cos}[\theta_1]\text{sin}[\theta_1]^2-\text{cos}[\theta_2]\text{sin}[\theta_2]^2)\cdot\frac{e^3}{h^2\omega},\label{eq:sigma_linear_node}\\
\mbox{and: }\theta_1 &\equiv\widetilde{\mbox{arccos}}\big[\frac{\frac{2\mu}{\hbar\omega}+1}{W}\big],\;\theta_2\equiv\widetilde{\mbox{arccos}}\big[\frac{\frac{2\mu}{\hbar\omega}-1}{W}\big].
\end{align}
Up to an order unity constant $\sigma^{abc}_{shift}(\omega)$ is determined by  $\sigma^{linear}_{node}(\omega)$. Here $W$ is defined in Eq.(\ref{eq:Type_II_criterion}), $\epsilon_{abc}$ is the Levi-Civita antisymmetric tensor, and we have chosen the convention $\omega>0$. $\chi\equiv \mbox{sign}[v^xv^yv^z]=\pm1$ is the chirality of the Weyl node (i.e., monopole charge). Note that we have chosen the frame $x,y,z$ to be right-handed. We have also defined a function $\widetilde{\mbox{arccos}}[s]\equiv0$ if $s\geqslant1$, $\widetilde{\mbox{arccos}}[s]\equiv\pi$ if $s\leqslant-1$, and $\widetilde{\mbox{arccos}}[s]\equiv\mbox{arccos}[s]$ if $-1<s<1$.

Interestingly, due to the chirality factor in this result, we know that this $1/\omega$ divergent term is absent in Dirac semimetals, because in those systems each Dirac node can be viewed as two Weyl nodes with opposite chiralities sharing the same set of $\mu$, $\vec v_t$ and opposite $\vec v$. It can also be immediately seen from Eq.~\eqref{eq:sigma_line_node} that the $1/\omega$ divergence is absent for  $\sigma^{aaa}_{shift}$ due to the Levi-Civita antisymmetric tensor.

The physical meaning of the two angles $\theta_1$,$\theta_2$ is the following. For the moment it is convenient to re-scale the momentum in $x,y,z$ directions so that $v^x=v^y=v^z=v$. The energy conservation $\delta(\omega_{mn}-\omega)$ constrains our consideration on a sphere in the re-scaled momentum space, whose radius equals $\frac{2\omega}{v}$. The re-scaled $\vec v_t$ can be used to define a special axis to set up a spherical coordinate system. Generally speaking, Pauli blocking takes over in certain solid angle regions. Namely the factor $f_{nm}$ further constrains the sphere into a region between the polar angles $\theta_1$ and $\theta_2$ (See Fig.\ref{fig:illustration}(e) for an illustration). 

In Type-I WSM ($W<1$) at zero doping, the whole sphere contributes, consistent with $\theta_1=0$ and $\theta_2=\pi$ (Fig.\ref{fig:illustration}(b)). But this is exactly a situation when the $\theta$-dependent factor in Eq.(\ref{eq:sigma_linear_node}), and thus the $1/\omega$ term in $\sigma_{shift}$, vanishes. (Note that, in this case, a careful analysis including band-bending effects shows that even the constant order vanishes, leaving the next order $\sigma_{shift}\propto\omega$ in the low frequency limit. See supplemental information.) But in Type-II WSM ($W>1$) at zero doping, only the part of the sphere between $\theta_{1,2}=
\mbox{arccos}[\pm\frac{1}{W}]$ contributes (Fig.\ref{fig:illustration}(e)), leading to the $1/\omega$ divergent response.

This results from the strongly angle dependent diverging Berry connection in WSM, despite the integrand $I_{mn}^{abc}$ in Eq.(\ref{eq:general_shift}) scaling as $\frac{1}{\omega^3}$. The full angular average over the $4\pi$ solid angle would annihilate the $1/\omega$ term in $\sigma_{shift}$. However, Pauli-blocking could take over in certain angular regions, removing the net cancellation and retaining the divergent term.

In particular, at $T=0$, in Type-I WSM ($W<1$) with finite $\mu$, this $1/\omega$ term survives only over a frequency range: $\frac{2|\mu|}{1+W}\leqslant\hbar\omega\leqslant\frac{2|\mu|}{1-W}$. (Thus in the extreme case when $v_t=0$ and $W=0$, this term vanishes.) In fact, it is straightforward to show that, $\sigma^{linear}_{node}$ would also change sign exactly at $\hbar\omega=2|\mu|$. Altogether this leads to a large peak in the $\sigma_{shift}(\omega)$ with a width $\sim |\mu|/\hbar$ and a height $\sim \frac{e^3}{h}\cdot\frac{1}{|\mu|}$, featuring a characteristic sign-changing line-shape (Fig.\ref{fig:line_shapes} top). Similarly a finite $\mu$ truncates the $1/\omega$ divergence in Type-II WSM and leads to a sign-changing large peak when $\hbar\omega\sim \mu$ (Fig.\ref{fig:line_shapes} bottom).

As shown in Fig.\ref{fig:line_shapes}, the line-shapes of these peaks are quasi-universal: they only depend on the dimensionless number $W$. Different values of doping $\mu$ only re-scale the peak's frequency range by a $\mu$ factor and its height by a $1/\mu$ factor. This means that one may use $\mu$ to control the frequency range of the peak to engineer tunable frequency-sensitive photo-electric devices.

\begin{widetext}

\begin{figure}[h]
\includegraphics[width=\textwidth]{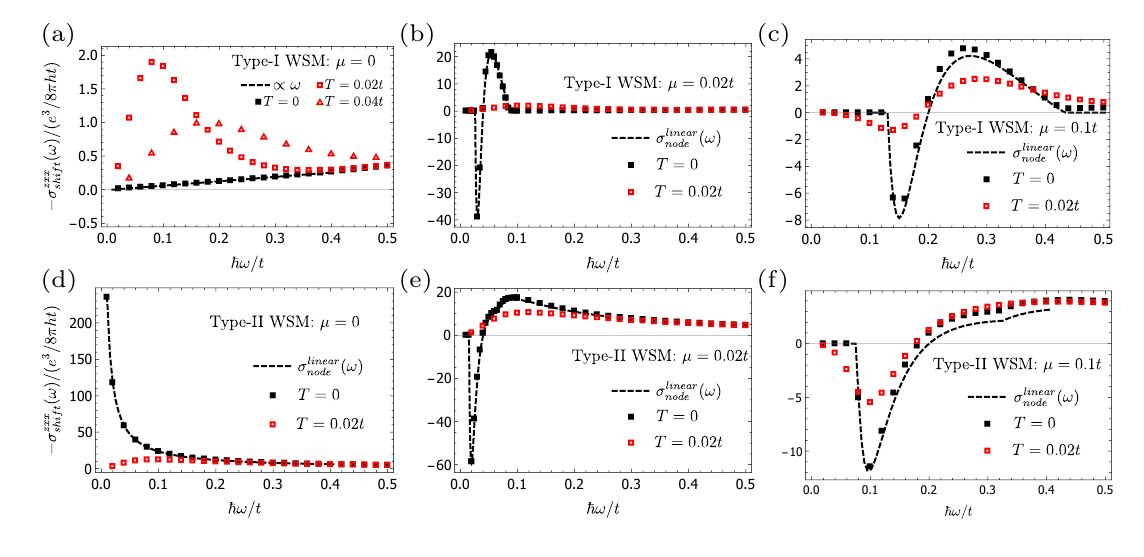}
\caption{(color online) Numerically computed $\sigma^{zxx}_{shift}(\omega)$ using the full tight-binding model Eq.(\ref{eq:tight_binding_model}) with parameters in the main text (squares and triangles), comparing with analytic linear-node results after summing over four Weyl nodes $\sigma^{linear}_{node}(\omega)$ (dashed lines). At zero doping, (a): $\sigma_{shift}\propto\omega$ at $T=0$ in Type-I WSM in the low frequency regime; a finite temperature partially plays the role of doping and induces a peak of $\sigma_{shift}$ whose width $\propto T$ and height $\propto 1/T$ (see supplemental material Fig.\ref{fig:temperature}); (d):  $\sigma_{shift}\propto1/\omega$ at $T=0$ in Type-II WSM, fully consistent with the result Eq.(\ref{eq:sigma_line_node}) within the linear approximation. This divergence is truncated by a finite temperature below $\hbar\omega\sim 5 k_B T$. (b)(c)(e)(f): At finite dopings $\sigma_{shift}$ feature large peaks whose width $\propto \mu$ and height $\propto 1/\mu$. At $T=0$ these large peaks are well captured by Eq.(\ref{eq:sigma_line_node}) (the slight deviations for $\mu=0.1t$ cases are due to expected band-bending effects.). At $k_BT=0.02t$ the peaks for $\mu=0.02t$ cases are strongly smeared out, while those for $\mu=0.1t$ are quantitatively reduced.}
\label{fig:tight_binding}
\end{figure}
\end{widetext}

\textit{Temperature and impurities---} At finite temperatures, the Fermi-Dirac distribution would smear out and truncate the divergences when $\hbar\omega\lesssim k_B T$ (which is confirmed in our tight-binding model calculations Fig.\ref{fig:tight_binding}). But at low temperatures $k_B T\ll \hbar\omega$ these divergences are not significantly modified. (See supplemental material Eq.(\ref{eq:universal_function},\ref{eq:general_sigma_with_universal_function}) for a general analytic form of the $\sigma_{shift}$ within linear-dispersion approximation.) However even at zero temperature, impurities give rise to scattering, while finite temperature enables other scattering mechanisms (e.g. electron-phonon). These scatterings, which can be phenomenologically characterized by a scattering time $\tau$, have been ignored so far. Namely, even at low temperatures $k_B T\ll \hbar\omega$ our result Eq.(\ref{eq:sigma_line_node}) holds only in the long scattering time $\omega\tau\gg1$ limit with the divergences truncated when $\omega\lesssim \frac{1}{\tau}$. For instance, previous experiments report that $\tau$ in the Type-I WSM TaAs is of the order of a pico-second\cite{parameswaran2014probing,behrends2016visualizing}. This suggests that our predicted striking response can be observed in the Terahertz or higher frequency regimes.

A conceptually interesting situation occurs when a finite temperature is introduced in the $\mu=0$ Type-I WSM, where divergence is absent at $T=0$ due to angular cancellation discussed before. However when $T\neq 0$, the thermally excited particle-hole pairs partially play the role of doping, and there is no reason for a full angular cancellation. In this case, simple dimensional analysis leads to striking results: we expect a temperature-induced peak of the intrinsic $\sigma_{shift}$ when $\hbar\omega\sim k_BT$, whose height $\sim \frac{e^3}{hk_B T}$ diverging in the low $T$ limit. (see supplemental material Eq.(\ref{eq:universal_function}) and Fig.\ref{fig:temperature}) This is observed in our tight-binding model calculations (see Fig.\ref{fig:tight_binding}(a)).

\textit{The size of the effects ---}It is interesting to estimate the size of effects in the divergent regimes; e.g., when $\hbar\omega\sim\mu$ for Type-I WSM or $\hbar\omega\gtrsim \mu$ in Type-II WSM, in the presence of a temperature $k_BT\ll \hbar\omega$. Previously shift-current responses have not be much studied in the low frequency (e.g. Terahertz) regimes. Note that even in a generic multiband metal, $\sigma_{shift}$ is expected to vanish at zero temperature in the low frequency regimes. This is simply because the energy conservation $\delta(\omega_{mn}-\omega)$ and $f_{nm}$ in $JDOS$ constrain both the valence band and the conduction band at the Fermi level when $\omega\rightarrow 0$, which would not occur due to band-repulsions.

Plugging in $\omega=1$THz, the estimated size of $\sigma_{shift}\sim \frac{e^3}{h^2}\frac{1}{\omega}$  in the divergent regimes is $\sim 0.01 A/V^2$, several orders of magnitudes larger than known reported values in visible or infrared regimes\cite{auston1972optical,glass1974high,koch1976anomalous,zenkevich2014PRB,somma2014high,zheng2014first}

\textit{Tight-binding model---} To concretely illustrate the predicted responses we compute the $\sigma_{shift}$ tensor using a minimal time-reversal symmetric 4-band tight-binding model featuring 4 Weyl nodes:\cite{wu2017NPhys}
\begin{align}
 &H^{TB}=t\big[(2.5 - \cos k^x - \cos k^y - \cos k^z) \tau_x +\sin k^y \tau_y \notag\\
 &+ 0.5 \cos k^y s_x\tau_y + \sin k^z s_x\tau_z + (\xi \cos k^x - \mu_0)\big]-\mu,\label{eq:tight_binding_model}
\end{align}
where $s_a$(spin) and $\tau_a$(orbital) are two sets of Pauli matrices. $t$ is an overall energy scale which can be $0.1-0.5$eV, to be broadly consistent with the relevant energy scales in existing Weyl materials. Symmetries in this model include: time-reversal $is_yK$ ($K$ is conjugation), $x\rightarrow -x$ mirror $s_x$, and $y\rightarrow -y$ mirror $s_y\tau_x$. The four Weyl nodes are located at $(k^x=\pm 0.920,k^y=\pm 0.464,k^z=0)$, with an energy $\mu_0t$ and $\mu_0=0.606\xi$. A positive parameter $\xi$ controls the tilting velocity $\vec v_t$: for $\xi<1$($\xi>1$), Type-I(Type-II) WSM is realized. We choose $\xi=0.5$ ($\xi=1.5$), corresponding to $W=0.539$($W=1.616$), as the representative for Type-I (Type-II) WSM.

We numerically compute $\sigma^{abc}_{shift}$, using the full formula Eq.(\ref{eq:general_shift}) without resorting to the linear approximation. The results of a particular component $\sigma^{zxx}_{shift}$ are plotted in Fig.\ref{fig:tight_binding} for various doping levels and temperatures, which are fully consistent with previous discussions.

\textit{Second harmonic generation---} Another directly related second order nonlinear optical response is the second harmonic generation (SHG), in which light at frequency $\omega$ drives current at frequency $2\omega$\cite{boyd2003nonlinear}. Defining SHG response tensor $\mathbf{j}^a(2\omega)=\sigma_{SHG}^{abc}(\omega,2\omega)E^b(\omega)E^c(\omega)$, it is known that the real part $\mbox{Re}[\sigma_{SHG}]$ is given by the interband contribution, and within the two-band approximation and linear-node approximation, we have
\begin{align}
    \mbox{Re}[\sigma^{abc}_{SHG}(\omega,2\omega)]=-\frac{3}{2}\sigma^{abc}_{shift}(\omega)\label{eq:SHG_shift_relation}
\end{align} 
where $\sigma^{abc}_{shift}(\omega)$ given by Eq.~\eqref{eq:sigma_line_node} (see supplemental information). \footnote{Previously another identity is known within general two-band approximation: \unexpanded{$\mbox{Re}[\sigma^{aaa}_{SHG}(\omega,2\omega)]=-\sigma^{aaa}_{shift}(\omega)+\frac{1}{2}\sigma^{aaa}_{shift}(2\omega)$} \cite{Nagaosa_sciadv.1501524}, which naively is inconsistent with Eq.(\ref{eq:SHG_shift_relation}). However within linear-node approximation we have shown that $\sigma^{aaa}_{shift}(\omega)=0$, which is consistent with both identities.} \footnote{Interestingly, the intraband contribution of SHG is responsible for \unexpanded{$\mbox{Im}[\sigma_{SHG}]$}, which has been pointed out to be a Fermi surface Berry's curvature effect and contains a $1/\omega$ term\cite{sodemann2015quantum,morimoto2016semiclassical}}. For example, similar to $\sigma_{shift}$, the zero temperature $\mbox{Re}[\sigma_{SHG}]$ features a similar $1/\omega$ divergence in Type-II WSM when $\mu=0$, and large peak behaviors in both Type-I and Type-II WSM when $\mu\neq0$. And even at $\mu=0$, a finite temperature induces a large peak in $\mbox{Re}[\sigma_{SHG}]$ in Type-I WSM with a width $\sim k_B T$ and a height $\sim \frac{e^3}{h k_B T}$.

\textit{Possible applications---} In this paper we report the $1/\omega$ diverging DC photovoltaic effect in the low frequency regime in WSM, due to the combination of the diverging Berry's curvature and Pauli-blocking effect. Fast Terahertz photon detection has been a long standing challenge. The reported large photovoltaic effect may be useful for this purpose. In addition, the doping-induced large peak regimes of $\sigma_{shift}$ in WSM, whose frequency ranges are controlled by $\mu$, may be useful as a tunable frequency-sensitive probe for far-infrared or Terahertz photons, i.e., a spectrum analyzer. 

XY and YR acknowledge support from the National Science Foundation under Grant No. DMR-1151440. KSB is grateful for support from the National Science Foundation, Grant No. DMR-1709987.

\bibliographystyle{apsrev}
\bibliography{nonlinear_optics_Weyl}

\appendix
\begin{widetext}
\begin{center}
\large{\bf Supplemental Material:\\ Divergent bulk photovoltaic effect in Weyl semimetals}\\
\vspace{14pt}
\normalsize{Xu Yang, Kenneth Burch, and Ying Ran}
\end{center}
\section{Shift current in type-I Weyl semi-metal}\label{app:Type_I}
In this section we shall prove that for a generic type-I Weyl semi-metal with Fermi level at the Weyl nodes, the leading term in the shift current tensor will be proportional to $\omega$ when $\omega\rightarrow 0$.

The low energy physics can be captured by the following generic 2-band Hamiltonian with chemical potential $\mu$ set to zero
\begin{equation}
    H=f_0\sigma_0+\sum\limits_{i=x,y,z}f_i\sigma_i,
\end{equation}
where $f_0,f_i$ are functions of $k$ and $f_i=0$ when $k=0$. The eigenvalues are $E_c=f_0+\epsilon$, $E_v=f_0-\epsilon$ with $\epsilon=\sqrt{\sum_i f_i^2}$. Since the tilting will not affect the shift current tensor, we will set $f_0$ to be zero.

The shift current tensor for the 2-band model is obtained by doing the following integral\cite{sipe2000second,2017NatCo...814176C}
\begin{equation}
 \sigma_{shift}^{abc}(\omega)=\frac{2\pi e^3}{\hbar^2}\int\frac{d^3\vec k}{(2\pi)^3}I_{cv}^{abc}[f_{cv}\cdot\delta(\omega_{cv}-\omega)],
\end{equation}
where $I^{abc}_{cv}$ has the following explicitly gauge-invariant expression
\begin{equation}
    I^{abc}_{cv}=\sum_{i,j,m}[\frac{1}{8\epsilon^3}(f_mf_{i,b}f_{j,ac}-f_{i,b}f_{j,a}f_m\frac{\epsilon_{,c}}{\epsilon})\epsilon_{ijm}+(b\leftrightarrow c)],
\end{equation}
where $\epsilon_{ijm}$ is the Levi-Civita symbol.

Now let's prove that terms proportional to $1/\omega$ (denoted as $\sigma^{(-1)}(\omega)$ below) and terms independent of $\omega$ (denoted as $\sigma^{(0)}(\omega)$ below) in $\sigma^{abc}$ vanish in the low frequency limit $\omega\rightarrow 0$. A simple order of estimate with $k\sim \omega$ when $\omega\rightarrow 0$ tells us that only the $k$-linear terms and $k$-quadratic terms in $f_i$ will contribute to $\sigma^{(-1)}(\omega)$ and $\sigma^{(0)}(\omega)$. Therefore it suffices to consider a linear node plus some quadratic corrections. It is always possible to choose the following form of $f_i$ by an affine transformation which does not affect the integral ($\hbar v_F$ is set to be 1 throughout this section)
\begin{equation}
    f_i=k_i+\alpha_{ijl}k_jk_l,
\end{equation}
where $\alpha_{ijl}$ is a rank-3 tensor symmetric respect to the interchange $j\leftrightarrow l$ and repeated indices are summed over.

In the spherical coordinate system, we have
\begin{equation}
    k_x=r \text{sin}[\theta]\text{cos}[\phi],k_y=r \text{sin}[\theta]\text{sin}[\phi],k_z=r \text{cos}[\theta].
\end{equation}
And the unit vector $\hat{k}=(\text{sin}[\theta]\text{cos}[\phi],\text{sin}[\theta]\text{sin}[\phi],\text{cos}[\theta])$ will be used below to simplify the notation.

When $\omega$ is small, $k_i\sim r\sim \omega$ and we can expand $\epsilon$ in powers of $r$
\begin{equation}
    \epsilon=r+r^2f(\theta,\phi)+\mathcal{O}(r^3),
\end{equation}
where $f(\theta,\phi)$ is a function of $\alpha_{ijl},\theta,\phi$. It is easy to obtain the explicit form of $f(\theta,\phi)$, but for our purpose we only need the fact that $f(\theta,\phi)=-f(\pi-\theta,\pi+\phi)$ since it comes from the angular dependence of terms involving 3 $k_i$'s.

In the same spirit we can expand $I^{abc}$ in powers of $r$
\begin{equation}
\begin{split}
&I^{abc}=\frac{1}{8\epsilon^3}[I^{(0)}_{abc}+I^{(1)}_{abc}+\mathcal{O}(r^2)],\text{ with }\\
&I^{(0)}=-\epsilon_{bam}\frac{k_mk_c}{\epsilon^2}+(b\leftrightarrow c)=-\epsilon_{bam}r^2\frac{\hat{k}_m\hat{k}_c}{\epsilon^2}+(b\leftrightarrow c),\\
&I^{(1)}=2\epsilon_{bjm}\alpha_{jac}k_m+\frac{h^{abc}(k_x,k_y,k_z)}{\epsilon^2}+(b\leftrightarrow c)
=2r\epsilon_{bjm}\alpha_{jac}\hat{k}_m+r^3\frac{h^{abc}(\hat{k}_x,\hat{k}_y,\hat{k}_z)}{\epsilon^2}+(b\leftrightarrow c)\end{split}
\end{equation}
where $h^{abc}(k_x,k_y,k_z)$ is a homogeneous polynomial of degree 3, whose explicit form, for our purpose, is not important.

The integral then becomes
\begin{equation}
    \sigma^{abc}(\omega)=const*\int d\Omega(\int r^2drI^{abc}(r,\theta,\phi)\delta(2\epsilon(r,\theta,\phi)-\omega)),
\end{equation}
where $d\Omega=\text{sin}[\theta]d\theta d\phi$.

When $\omega\rightarrow 0$, we know that for any fixed $\theta,\phi$ there is only one solution $r(\theta,\phi)$ to the equation $2\epsilon(r(\theta,\phi),\theta,\phi)=\omega$, from which we can solve $r$ as a function of $\theta$ and $\phi$. In fact, we can expand $r(\theta,\phi)$ in powers of $\omega$ with $f(\theta,\phi)$ defined before
\begin{equation}
    r(\theta,\phi)=\frac{\omega}{2}-\frac{\omega^2}{4}f(\theta,\phi)+\mathcal{O}(\omega^3).
\end{equation}

The integral can then be written as
\begin{equation}\label{eq:Type_I_shifttensor}
\begin{split}
&\sigma^{abc}(\omega)=const*\int d\Omega \frac{1}{\omega^3}\frac{ r(\theta,\phi)^2}{\partial_{r}\epsilon|_{r=r(\theta,\phi)}}[I^{(0)}_{abc}+I^{(1)}_{abc}+\mathcal{O}(\omega^2)]\\
&=const*\int d\Omega \frac{1}{\omega^3}[ r(\theta,\phi)^2-2r(\theta,\phi)^3f(\theta,\phi)+\mathcal{O}(\omega^4)][I^{(0)}_{abc}+I^{(1)}_{abc}+\mathcal{O}(\omega^2)]\\
&=const*\int d\Omega \frac{1}{\omega}[ 1-2\omega f(\theta,\phi)+\mathcal{O}(\omega^2)][I^{(0)}_{abc}+I^{(1)}_{abc}+\mathcal{O}(\omega^2)]
\end{split}
\end{equation}

Terms that are of order $1/\omega$ comes from the integral over $I^{(0)}_{abc}$ with $r(\theta,\phi)$ set to $\omega/2$ in Eq.\eqref{eq:Type_I_shifttensor}
\begin{equation}
\sigma^{(-1)}(\omega)=const*\int d\Omega \frac{1}{\omega}(I^{(0)}_{abc})=const*\int d\Omega \frac{1}{\omega}[-\epsilon_{bam}(\hat{k})_m(\hat{k})_c+(b\leftrightarrow c)]=const*(-\epsilon_{bac}-\epsilon_{cab})=0,
\end{equation}
where we have used the fact that $\int (\hat{k})_m(\hat{k})_nd\Omega=\frac{4\pi}{3}\delta_{m,n}$.

Terms that are independent of $\omega$ comes from the following 3 integrals:
\begin{equation}\label{eq:constantI}
    \sigma^{(0)}_{I}=const*\int d\Omega \frac{1}{\omega}(I^{(0)}_{abc})=const*\int d\Omega f(\theta,\phi)[-\epsilon_{bam}\hat{k}_m\hat{k}_c+(b \leftrightarrow c)],
\end{equation}
\begin{equation}\label{eq:constantII}
    \sigma^{(0)}_{II}=const*\int d\Omega \frac{1}{\omega}(I^{(1)}_{abc})=const*\int d\Omega [\epsilon_{bjm}\alpha_{jac}\hat{k}_m+h^{abc}(\hat{k}_x,\hat{k}_y,\hat{k}_z)/2+(b\leftrightarrow c)]
\end{equation}
and 
\begin{equation}\label{eq:constantIII}
    \sigma^{(0)}_{III}=const*\int d\Omega f(\theta,\phi)(I^{(0)}_{abc})=const*\int d\Omega f(\theta,\phi)[-\epsilon_{bam}(\hat{k})_m(\hat{k})_c+(b\leftrightarrow c)].
\end{equation}

It is easy to see that under $(\theta,\phi)\rightarrow (\pi-\theta,\pi+\phi)$, the integrand of Eq.\eqref{eq:constantI},\eqref{eq:constantII},\eqref{eq:constantIII} all change sign, therefore they are all equal to zero.

In conclusion, we have analytically shown that in the low frequency limit $\omega\rightarrow 0$, term that diverges as $1/\omega$ and term that is independent of $\omega$ in the shift current tensor $\sigma^{abc}$ vanish. Therefore the leading term in $\sigma^{abc}$ will be proportional to $\omega$.

\section{Analytical formula for the shift current in Weyl semi-metal with tilting and doping in low-frequency limit}\label{app:tilting}
In this section we will obtain an analytical formula of the shift-current tensor for a Weyl node with both tilting and nonzero chemical potential within the linear approximation at zero temperature. This formula captures the physics of type-I and type-II Weyl semimetals with or without doping in a unified fashion.

Let's consider the following generic Hamiltonian
\begin{equation}
H=(\hbar \vec{k}\cdot \vec{v}^{t}-\mu)\sigma_0+\sum\limits_{i=x,y,z} \hbar k_iv_i\sigma_i.
\end{equation}

We will choose the following parameterization $v_ik_i=rO_{ij}\hat{k}_j$, where $O_{ij}$ is an orthogonal matrix and $\hat{k}=(\text{sin}[\theta]\text{cos}[\phi],\text{sin}[\theta]\text{sin}[\phi],\text{cos}[\theta])$. The orghogonal matrix $O$ is chosen such that
\begin{equation}(\frac{v^t_{x}}{v_x},\frac{v^t_y}{v_y},\frac{v^t_z}{v_z})\cdot O=(0,0,W),
\end{equation} where $W=\sqrt{(\frac{v^t_{x}}{v_x})^2+(\frac{v^t_y}{v_y})^2+(\frac{v^t_z}{v_z})^2}$. In fact the third column of $O$ is fully determined\begin{equation}\label{eq:Omatrix}
    (O_{13},O_{23},O_{33})=\frac{1}{W}(\frac{v^t_{x}}{v_x},\frac{v^t_y}{v_y},\frac{v^t_z}{v_z}).
\end{equation}

In this new coordinate system, we have
\begin{equation}
     \sigma_{shift}^{abc}(\omega)=\frac{2\pi e^3}{\hbar^2}\int\frac{d^3\vec k}{(2\pi)^3}I_{12}^{abc}[f_{21}\cdot\delta(\omega_{12}-\omega)]=\frac{ e^3\text{sign}[v_xv_yv_z]}{4\pi^2\hbar^2v_xv_yv_z}\int r^2dr d\Omega I_{12}^{abc}f_{21}\cdot\delta(2r-\omega).
\end{equation}
where $I^{abc}_{12}$ is given by
\begin{equation}
    -\frac{v_av_bv_c}{8r^3}[O_{mj}\hat{k}_jO_{cl}\hat{k}_l\epsilon_{bam}+(b\leftrightarrow c)]
\end{equation}
and $f_{21}$ is given by
\begin{equation}
    f_{21}=\frac{1}{e^{\beta (\hbar Wr\text{cos}[\theta]-\hbar r-\mu)}+1}-\frac{1}{e^{\beta (\hbar Wr\text{cos}[\theta]+\hbar r-\mu)}+1},
\end{equation}
which is a function of $r$ and $\theta$.

Therefore we have
\begin{equation}\label{eq:shifttensorintegral}
    \sigma_{shift}^{abc}(\omega)=-\frac{ e^3\text{sign}(v_xv_yv_z)v_av_bv_c}{8\omega h^2v_xv_yv_z}\int \text{sin}[\theta]d\theta d\phi f_{21}(\omega/2,\theta) [O_{mj}\hat{k}_jO_{cl}\hat{k}_l\epsilon_{bam}+(b\leftrightarrow c)].
\end{equation}

The angular integration can be easily done: 
\begin{equation}\label{eq:angularintegration}
    \int d\Omega f_{21}(\omega/2,\theta)\hat{k}_m\hat{k}_n=\int d\Omega f_{21}(\omega/2,\theta)(\hat{k}_x^2 \delta_{m,n}+(\hat{k}_z^2-\hat{k}_x^2)\delta_{m,n}\delta_{m,3})=c\cdot \delta_{m,n}-\pi\cdot g(\varpi,z,W)\delta_{m,n}\delta_{m,3},
\end{equation}
where $c$ is a constant, $z=e^{\beta \mu}$ is the fugacity and $\varpi=\frac{\beta\hbar\omega}{2}$. The isotropic part $c\cdot \delta_{m,n}$ does not contribute to the shift current tensor as in the case of type-I Weyl semimetal without doping. And the anisotropic part can be evaluated by first integrating over $\phi$ and then integrating over $x=\text{cos}[\theta]$ to yield
\begin{equation}
\label{eq:universal_function}
\begin{split}
    &g(\varpi,z,W)=-\int^{1}_{-1}(3x^2-1)f_{21}(\omega/2,x)dx=\frac{2}{W\varpi}\text{ln}[\frac{1+z^{-1}e^{(1-W)\varpi}}{1+z^{-1}e^{(1+W)\varpi}}\cdot \frac{1+z^{-1}e^{(W-1)\varpi}}{1+z^{-1}e^{-(1+W)\varpi}}]\\
    &-\sum\limits_{n=2,3}\frac{6}{W^n\varpi^n}[\text{Li}_n(-z^{-1}e^{(1-W)\varpi})+\text{Li}_n(-z^{-1}e^{(1+W)\varpi})-\text{Li}_n(-z^{-1}e^{-(1+W)\varpi})-\text{Li}_n(-z^{-1}e^{(-1+W)\varpi})],
    \end{split}
\end{equation}
where $\text{Li}_n(x)$ is the polylogarithm of order $n$.

Therefore after inserting Eq.\eqref{eq:Omatrix} and Eq.\eqref{eq:angularintegration} into Eq.\eqref{eq:shifttensorintegral} we have the following result
\begin{align}\label{eq:general_sigma_with_universal_function}
  \sigma^{abc}_{shift}(\omega)&=\frac{\pi\text{sign}(v_xv_yv_z)}{8W^2}\cdot \big[\epsilon_{dca}\frac{v^t_bv^t_d}{(v_d)^2}+(b\leftrightarrow c)\big]\cdot\sigma^{linear}_{node}(\omega)\\
  &\mbox{where: }\sigma^{linear}_{node}(\omega)\equiv\big[g(\varpi,z,W)\cdot\frac{e^3}{h^2\omega}\big]\notag
\end{align}

\begin{figure}
    \centering
    \includegraphics[width=0.4\textwidth]{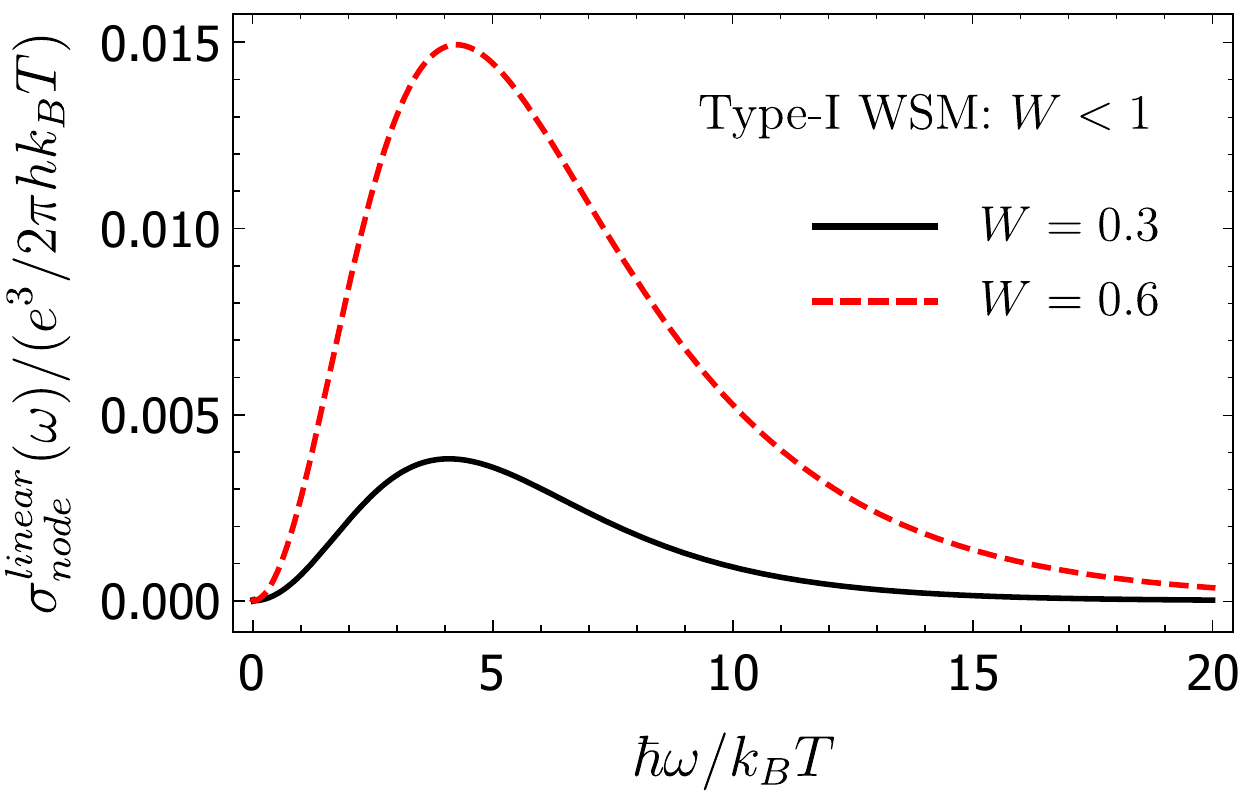}\;\;\includegraphics[width=0.4\textwidth]{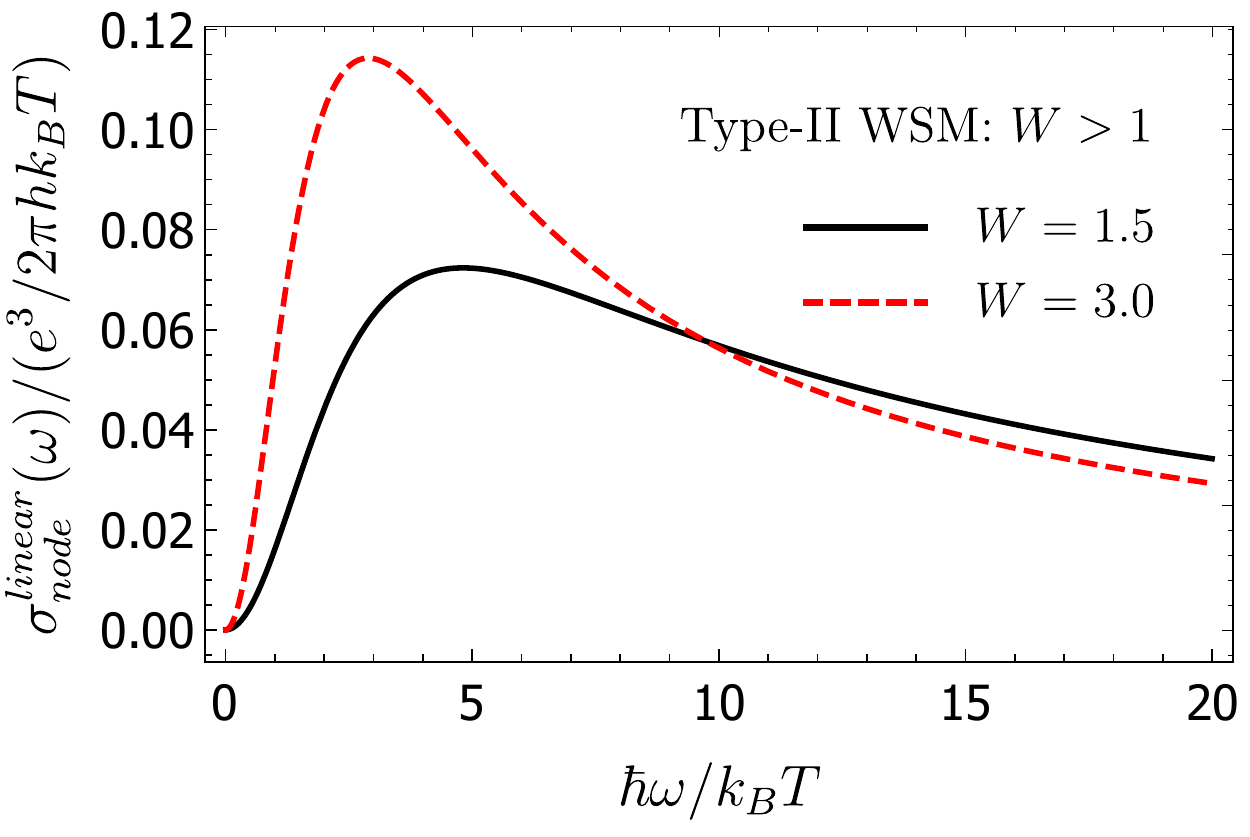}
    \caption{At zero doping $\mu=0$, based on the linear-node result Eq.(\ref{eq:universal_function}), we find that a finite temperature induces a peak of $\sigma_{shift}$ in Type-I WSM (left), and truncate the $1/\omega$ divergence in Type-II WSM(right) when $\hbar\omega\sim k_BT$. Note that the frequency range is re-scaled by a $k_BT$ factor while $\sigma_{shift}$ is re-scaled by a $1/k_BT$ factor. The line shapes of these curves only depend on $W$ but independent of $k_BT$.}
    \label{fig:temperature}
\end{figure}

Therefore up to an order 1 constant, the shift-current response tensor is determined by the function $\sigma^{linear}_{node}$. Let's discuss two simple limits of Eq.~\eqref{eq:universal_function}. First, when $\mu=0$, the function $g$ only depends on the dimensionless variable $\varpi$, which gives us the scaling form \begin{equation}
    \sigma_{shift}(\omega)|_{\beta}=b\cdot\sigma_{shift}(b\omega)|_{\beta/b}
\end{equation} 
We plot this scaling behavior at finite temperatures and $\mu=0$ for a few representative values of $W$ in Fig.\ref{fig:temperature}. In particular for Type-I WSM at $\mu=0$, $T\neq 0$ induces a peak of $\sigma_{shift}$ whose width$\propto k_BT$ and height$\propto 1/k_B T$. This behavior is also observed in Fig.3(a) in our tight-binding model calculations.

Next, when $\beta\rightarrow \infty$, we can obtain a simpler form of $g(\varpi,z,W)$. In fact, an easier way is to replace the Fermi-Dirac distribution function by the Heaviside step function: \begin{equation}
    f_{21}=\Theta(r-\vec{k}\cdot \vec{v}^t+\mu/\hbar)-\Theta(-r-\vec{k}\cdot \vec{v}^t+\mu/\hbar)=\Theta(r(1-W \text{cos}[\theta])+\mu/\hbar)-\Theta(-r(1+W\text{cos}[\theta])+\mu/\hbar).
\end{equation}

The Heaviside step function together with the $\delta$ function constraint the integration region of $\theta$, therefore we will introduce the following two $\theta$ angles to characterize the upper and lower limit of the integration
\begin{equation}
    \theta_1 \equiv\widetilde{\mbox{arccos}}\big[\frac{\frac{2\mu}{\hbar\omega}+1}{W}\big],\;\theta_2\equiv\widetilde{\mbox{arccos}}\big[\frac{\frac{2\mu}{\hbar\omega}-1}{W}\big]
\end{equation}
where the function $\widetilde{\mbox{arccos}}[s]$ is defined in the following way: $\widetilde{\mbox{arccos}}[s]\equiv0$ if $s\geqslant1$, $\widetilde{\mbox{arccos}}[s]\equiv\pi$ if $s\leqslant-1$, and $\widetilde{\mbox{arccos}}[s]\equiv\mbox{arccos}[s]$ if $-1<s<1$.

After integration we have
\begin{equation}
\sigma^{linear}_{node}(\varpi,z,W)|_{T\rightarrow 0}=(\text{cos}[\theta_1]\text{sin}[\theta_1]^2-\text{cos}[\theta_2]\text{sin}[\theta_2]^2)\frac{e^3}{h^2\omega},
\end{equation}
which gives us the shift-current tensor at zero-temperature:
\begin{equation}
  \sigma_{shift}^{abc}(\omega)=\frac{\pi\text{sign}(v_xv_yv_z)}{8W^2}\cdot \big[\epsilon_{dca}\frac{v^t_bv^t_d}{(v_d)^2}+(b\leftrightarrow c)\big]\cdot (\text{cos}[\theta_1]\text{sin}[\theta_1]^2-\text{cos}[\theta_2]\text{sin}[\theta_2]^2)\frac{e^3}{h^2\omega}
\end{equation}

\section{Analytical formula for the second-harmonic-generation in Weyl semi-metal with tilting and doping in low-frequency limit}\label{app:SHG}
The second-harmonic-generation (SHG) response tensor is defined via $\mathbf{j}^a(2\omega)=\sigma_{SHG}^{abc}(\omega,2\omega)E^b(\omega)E^c(\omega)$.

From the standard time-dependent perturbation theory\cite{boyd2003nonlinear,moss1990band,ghahramani1991full}, we have the following expression for the real part of the SHG response tensor
\begin{equation}\label{eq:SHG_multiband}
\begin{split}
&\text{Re}[\sigma_{SHG}^{abc}(\omega,2\omega)]=\frac{i\pi e^3}{2\hbar^2\omega^2}\sum\limits_{m,n,p}\int_{BZ}\frac{dk^3}{(2\pi)^3}[(v^a_{mn}w^{bc}_{nm}+\frac{2v^a_{mn}\{v^b_{np}v^c_{pm}\}}{\omega_{mp}+\omega_{np}})f_{mn}\delta(2\omega-\omega_{nm})\\
&+(w^{ab}_{mn}v^{c}_{nm}+w^{ac}_{mn}v^{b}_{nm})f_{mn}\delta(\omega-\omega_{nm})+\frac{v^a_{mn}\{v^b_{np}v^c_{pm}\}}{\omega_{pm}+\omega_{pn}}(f_{mp}\delta(\omega-\omega_{pm})-f_{np}\delta(\omega-\omega_{np}))],
\end{split}
\end{equation}
where $v^i_{mn}\equiv \frac{1}{\hbar}\bra{m}\partial_{k_i}H\ket{n}$, $w^{ij}_{mn}\equiv \frac{1}{\hbar^2}\bra{m}\partial_{k_i}\partial_{k_j}H\ket{n}$ and $\{v^b_{np}v^c_{pm}\}=v^b_{np}v^c_{pm}+v^c_{np}v^b_{pm}$. Note that since we are dealing with generic tight-binding models, a careful derivation following Ref.\cite{moss1990band} yields the extra $w^{ij}_{mn}$ terms which are absent in the literatures listed above.

For a 2-band model within the linear approximation, Eq.~\eqref{eq:SHG_multiband} can be simplified to be
\begin{equation}\begin{split}\label{eq:SHG_2_band}
&\text{Re}[\sigma_{SHG}^{abc}(\omega,2\omega)]=\frac{i\pi e^3}{2\hbar^2}\int_{BZ}\frac{dk^3}{(2\pi)^3}\frac{v^a_{vc}\{v^b_{cv}\Delta^c_{cv}\}}{\omega_{cv}^3}f_{cv}[8\delta(2\omega-\omega_{cv})-\delta(\omega-\omega_{cv})],
\end{split}
\end{equation}
where $\Delta^i_{nm}\equiv v^i_{nn}-v^i_{mm}$ and $v/c$ corresponds to valence/conduction bands respectively.

In the linear approximation, we have the following identity
\begin{equation}
    r^b_{mn}r^c_{nm;a}+r^c_{mn}r^b_{nm;a}=-\frac{v^a_{nm}\{v^b_{mn}\Delta^c_{mn}\}}{\omega_{mn}^3}-\frac{\Delta^a_{mn}\{v^b_{mn}v^c_{nm}\}}{\omega_{mn}^3}.
\end{equation}

Therefore we can rewrite Eq.~\eqref{eq:SHG_2_band} as follows
\begin{equation}\label{eq:SHG_shift}
    \text{Re}[\sigma_{SHG}^{abc}(\omega,2\omega)]=\frac{i\pi e^3}{2\hbar^2}\int_{BZ}\frac{dk^3}{(2\pi)^3}f_{vc}(r^b_{cv}r^c_{vc;a}+r^c_{cv}r^b_{vc;a})[8\delta(2\omega-\omega_{cv})-\delta(\omega-\omega_{cv})],
\end{equation}
where we have used the fact that the integration over $\frac{\Delta^a_{cv}\{v^b_{cv}v^c_{vc}\}}{\omega_{cv}^3}$ vanishes due to time-reversal symmetry.

Eq.~\eqref{eq:SHG_shift} assumes a very similar form to the expression of shift-current. In fact, it's easy to see that within the linear approximation we have
$\text{Re}[\sigma_{SHG}^{abc}(\omega,2\omega)]=-4\sigma^{abc}_{shift}(2\omega)+\frac{1}{2}\sigma^{abc}_{shift}(\omega)=-\frac{3}{2}\sigma^{abc}_{shift}(\omega)$, where $\sigma^{abc}_{shift}(\omega)$ is given by Eq.~\eqref{eq:general_sigma_with_universal_function}. As a consequence, all the discussions of leading-order terms in $$\text{Re}[\sigma_{SHG}^{abc}(\omega,2\omega)]$$ with the presence of tilting and doping at non-zero temperature naturally follow that of $\sigma^{abc}_{shift}(\omega)$ in the last section.

\end{widetext}

\end{document}